%% file: main.tex
\documentclass[runningheads,table,xcdraw]{llncs}
\input{addon/pkgloading}

\input{addon/glossary}
\usepackage{tabularx}
\usepackage{longtable}
\usepackage{cals}
\usepackage{ltablex}
\usepackage{booktabs}

\newcommand\pattern[1]{{\textsl{\gls{#1}}}}

\newcommand\mysubsubsection[1]{\medskip\noindent\textbf{#1.}}

\begin{document}

\title{Foundational Oracle Patterns: \\ Connecting Blockchain to the Off-chain World} 
\titlerunning{Foundational Blockchain Oracle Patterns}

\author{
    Roman~Mühlberger\inst{1} \and
    Stefan~Bachhofner\inst{1} \and
    Eduardo~Castell{\'o}~Ferrer\inst{2} \and
    Claudio~Di~Ciccio\inst{3} \and
    Ingo~Weber\inst{4} \and
    Maximilian~Wöhrer\inst{5} \and
    Uwe~Zdun\inst{5}
}

\authorrunning{Mühlberger, Bachhofner, Castell{\'o}~Ferrer, Di~Ciccio, Weber, Wöhrer, Zdun}

\institute{
    Vienna University of Economics and Business, 
    Vienna, Austria \\
    \and
    Massachusetts Institute of Technology,
    Cambridge, United States \\
    \and
    Sapienza University of Rome, 
    Rome, Italy \\
    \and
    Technische Universitaet Berlin, Berlin, Germany
    \and
    University of Vienna, Vienna, Austria
}
\maketitle 
\begin{abstract}
Blockchain has evolved into a platform for decentralized applications, with beneficial properties like high integrity, transparency, and resilience against censorship and tampering. 
However, blockchains are closed-world systems which do not have access to external state.
To overcome this limitation, oracles have been introduced in various forms and for different purposes.
However so far common oracle best practices have not been dissected, classified, and studied in their fundamental aspects.
In this paper, we address this gap by studying foundational blockchain oracle patterns in two foundational dimensions  characterising the oracles: (i) the data flow direction, i.e., inbound and outbound data flow, from the viewpoint of the blockchain; and (ii) the initiator of the data flow, i.e., whether it is push or pull-based communication.
We provide a structured description of the four patterns in detail, and discuss an implementation of these patterns based on use cases.
On this basis we conduct a quantitative analysis, which results in the insight 
that the four different patterns are characterized by distinct performance and costs profiles. %, and sheds light on these profiles.

\end{abstract}

\section{Introduction}
\label{sec:introduction}
\input{sections/introduction}

\section{Background and state of the art}
\label{sec:background}
\input{sections/background.tex}

\section{Patterns}
\label{sec:patterns}
\input{sections/pattern.tex}

\section{Use cases}
\label{sec:implementation}
\input{sections/implementation.tex}

%\section{Evaluation}
\section{Analysis of Performance and Transaction Fees}
\label{sec:evaluation}
\input{sections/evaluation.tex}

\section{Discussion and Threats to Validity}
\label{sec:discussion}
\input{sections/discussion.tex}

\section{Conclusion}
\label{sec:conclusion}
\input{sections/conclusion.tex}

%\section{Acknowledgements}
%\label{sec:acknowledgements}
\input{sections/acknowledgements.tex}

% \input{sections/appendix}

\bibliographystyle{splncs04}
\bibliography{bibliography}

\end{document}

%% file: addon/pkgloading.tex
\usepackage[utf8]{inputenc}
\usepackage{graphicx}
\usepackage{xcolor}
\usepackage[textsize=scriptsize,backgroundcolor=yellow!40]{todonotes}
\usepackage{paralist}
\usepackage{bera}

\usepackage{array}
\usepackage{booktabs}
\usepackage{multicol}
\usepackage{multirow}
\usepackage{rotating}
\usepackage{float}
\usepackage{comment}

\usepackage{footmisc}

\usepackage[nonumberlist,acronym,sanitize=none]{glossaries}

\usepackage[pdftex, colorlinks=true, hyperfootnotes=true, hyperindex=true,
            plainpages=false, pagebackref=false, pdfpagelabels=true, pdfstartview=FitH,
            linkcolor=blue, citecolor=blue, urlcolor=blue, breaklinks=true,
            bookmarks, bookmarksopen, bookmarksdepth=3]{hyperref}

\usepackage{times}
\usepackage[subtle]{savetrees}
\usepackage[capitalise,nameinlink]{cleveref}

\renewcommand{\arraystretch}{1.4}

%% file: addon/glossary.tex
\newacronym{iot}{IoT}{Internet of Things}
\newacronym{bpm}{BPM}{Business Process Management}

\newglossaryentry{pushinbo}{
	name={push-based inbound oracle},description={An oracle pushing off-chain information on-chain}}
\newglossaryentry{pushoutbo}{
	name={push-based outbound oracle},description={An oracle pushing on-chain information off-chain}}
\newglossaryentry{pullinbo}{
	name={pull-based inbound oracle},description={An oracle pulling off-chain information on-chain}}
\newglossaryentry{pulloutbo}{
	name={pull-based outbound oracle},description={An oracle pulling on-chain information off-chain}}

%% file: sections/introduction.tex
Conceptually, a blockchain is an append-only store for transactions, which is distributed across many machines and structured into a linked list of blocks~\cite{Xu.etal/2019:ArchitectureforBlockchainApplications}.
Based on its decentralized nature, structure, and use of cryptographic protocols, blockchain technology provides a modern platform for distributed applications with properties like high integrity, transparency, and resilience against censorship and tampering.
This creates, among others, new opportunities and challenges for inter-organizational business processes~\cite{DBLP:journals/tmis/MendlingWABCDDC18}. 
These inherent properties make blockchain technology a good fit for use cases where data integrity is of crucial importance, e.g. clinical trials~\cite{wong2019prototype,glicksberg2020blockchain},
food security~\cite{ahmed2017blockchain}, or financial risk when dealing with business partners~\cite[Ch.12]{Xu.etal/2019:ArchitectureforBlockchainApplications}.
Consequently, organizations realize efficiency and effectiveness gains with blockchain technology as business processes can have a higher degree of automation, e.g., by running business processes on the blockchain~\cite{DBLP:conf/bpm/WeberXRGPM16} or by automating information exchange between mutually untrusting parties. 
Many such applications are made possible by a feature of second-generation blockchains, \emph{smart contracts}, which ``are programs deployed as data in the blockchain ledger, and executed in transactions on the blockchain''~\cite{Xu.etal/2019:ArchitectureforBlockchainApplications}. 
With smart contracts, blockchains become decentralized, neutral execution platforms for user code.

Regardless of the generation, blockchains are closed-world systems: from inside, one can only access data that is on the blockchain already. 
Oracles have been proposed to mitigate that limitation. 
In the context of blockchains, an oracle is a component that can transfer data between the outside world and the blockchain.
However, the implementation of oracles provides considerable conceptual challenges as they can be regarded as a centralized point of failure or may introduce security and trust concerns~\cite{DBLP:journals/tmis/MendlingWABCDDC18}. 
Consequently, much of the research regarding oracles focuses on how to address these security and trust concerns, e.g., by using multiple independent oracle instances to form a decentralized oracle~\cite{Xu.etal/EuroPLoP2018:PatternCollectionBlockchainBasedApplications}, extending trust properties to off-chain computation~\cite{eberhardt2017offchaining_patterns}, or strengthening trust in incoming data~\cite{heiss2019trustworthyOnchaining}. % teutsch2017decentralized, 
However, foundational aspects of blockchain oracles that allow for their categorization and abstraction have not been subject to close investigation yet.

\begin{table}[t]
    \caption{An overview of the four oracle types.}
	\label{tab:oracle:types}
    \centering
    \setlength{\tabcolsep}{7pt}
    \renewcommand{\arraystretch}{1.5}
    \resizebox{\textwidth}{!}{%
    	\input{tables/oracle-types}
	}
\end{table}

In this paper, we address this gap by examining two core dimensions of oracles:  
(i) the \emph{direction}, i.e., whether the data flow is \emph{inbound} or \emph{outbound} from the viewpoint of the blockchain; and 
(ii) the \emph{initiator} of the data flow, i.e., whether it is \emph{push} or \emph{pull}-based communication.
There are four combinations of these options, an overview of which is shown in \cref{tab:oracle:types}.
We describe each of these as a pattern, and examine its characteristics. 
Note that, on this level, the four patterns can be implemented without relying on smart contracts, i.e., even on first-generation blockchains like Bitcoin.
Each of the patterns can also be suitably combined with other, higher-level patterns from the literature, like decentralization or provable computation. 

To characterise the different patterns, we implemented them in the context of two use cases, and use these implementations for the purpose of obtaining measurements.
To this end, the implementations are based on Ethereum, and we sent over $2,500$ transactions to the Ethereum test network to obtain concrete data.
This allows us to quantitatively study the characteristic differences between the four oracle patterns.
In particular, we focus on time (latency) and cost. 

\setcounter{footnote}{0}
% Action
The remainder of the paper is structured as follows. 
\cref{sec:background} introduces background literature and related work.
The patterns are described and contrasted in \cref{sec:patterns}.
The use cases for the implementation are described in \cref{sec:implementation}. On the basis of the implementation, we analyze the four patterns with respect to time and costs in \cref{sec:evaluation}.%
\footnote{The source code can be found at \url{https://github.com/MacOS/blockchain-oracles-data-collection} \label{fn:code-repo}}
Next, we discuss our results and threats to validity in \cref{sec:discussion}.
Finally, the paper concludes in \cref{sec:conclusion}.

%% file: tables/oracle-types.tex
\sffamily
\begin{tabular}{ r c | c }
	\cellcolor{gray!30}                  & \cellcolor{gray!30}\textbf{Pull}                                                                                       & \cellcolor{gray!30}\textbf{Push} \\
	\cellcolor{gray!30}\textbf{Inbound}  & \begin{tabular}[c]{c} The \textbf{on-chain} component \textbf{requests} the off-chain state \\ from an \textbf{off-chain} component \end{tabular} & \begin{tabular}[c]{c}The \textbf{off-chain} component \textbf{sends} the off-chain state\\ to the \textbf{on-chain} component\end{tabular} \\ \hline
	\cellcolor{gray!30}\textbf{Outbound} & \begin{tabular}[c]{c}The \textbf{off-chain} component \textbf{retrieves} the on-chain state\\ from an \textbf{on-chain} component\end{tabular}    & \begin{tabular}[c]{c}The \textbf{on-chain} component \textbf{sends} the off-chain state\\ to an \textbf{off-chain} component\end{tabular} \\
\end{tabular}%

%% file: sections/background.tex
In a significant number of times, 
applications built on blockchain infrastructure require data from real world states and events~\cite{beniiche2020study,DBLP:conf/caise/CiccioMP20}. 
Examples include financial data, weather-related information, random number generations or arbitrary data from off-chain devices and web services accessible via Application Programming Interfaces (APIs). 
Blockchain oracles provide a way to interact with the off-chain world~\cite{Xu.etal/2019:ArchitectureforBlockchainApplications}.
Oracles can be implemented as software (interacting with online sources) or hardware (interacting with the physical world), human (interacting with individuals) or computation-based oracles (performing off-chain calculations), single-source (centralized) or consensus-based oracles (decentralized, using a multitude of sources)~\cite{beniiche2020study}.
In this paper, we abstract from the way in which oracles are implemented and focus on the foundational patterns they realize. Next, we discuss the basic notions behind blockchains and elaborate on state-of-the-art solutions adopted thus far for the realization of oracles.

\mysubsubsection{Blockchain}
% The first-generation blockchain platforms such as Bitcoin~\cite{nakamoto2008bitcoin} are digital cash systems that operate without central authorities acting as an intermediary.
%
At the core of a blockchain lies the transaction, that is, the transfer of value between accounts.
Transactions are temporally ordered and stored in a sequential structure named ledger.
Every participating full node in the blockchain network keeps a local copy of the ledger.
Updates in the network are communicated via blocks, each collating the transactions to be appended to the ledger.
To generate and broadcast new blocks, the so-called mining nodes can be required to prove their trustworthiness, e.g., by solving computationally hard problems (Proof of Work).
A consensus algorithm allows for the eventual consistency of the distributed ledger.
Every block is linked to the previous one via hashing, thus forming a chain -- hence the name, blockchain.
% After several derivatives of first generation, Buterin~\cite{buterin2014next} introduced a new generation of blockchains: Ethereum.
% Among other improvements, Ethereum extended the concept of blockchain so as to turn it into a programmable infrastructure through the introduction of smart contracts, as introduced above.
Smart contracts turn blockchains such as Ethereum~\cite{wood2014ethereum}, Hyperledger Fabric~\cite{dhillon2017hyperledger} and Algorand~\cite{Chen.Micali/TCS2019:Algorand} into programmable infrastructures.
Developers can encode smart contracts with a programming language % such as Solidity%
%\footnote{Solidity project: \url{https://github.com/ethereum/solidity}. Accessed: June 7, 2020.} 
and compile them to bytecode.
Upon deployment, smart contracts are associated with a unique address.
They are executed and saved across all connected nodes of the network.
The invocations have a computation price expressed in terms of gas.
% The Ethereum Virtual Machine executes the smart contract bytecode, 
% which can be referred to via a unique address associated to an account for each individual, deployed smart contract. 
In order to store information, e.g., on the Ethereum blockchain, it can be placed into a transaction payload and possibly added to the contract storage, contract logs, or kept in the transaction payload~\cite{DBLP:conf/wicsa/XuPZGPTC16}. After the transaction is included into a block, the information is publicly accessible within the network.
% Smart contracts also exist in various other distributed ledger technologies, including Hyperledger Fabric~\cite{dhillon2017hyperledger} and Algorand~\cite{Chen.Micali/TCS2019:Algorand}. 

\mysubsubsection{Blockchain oracles}
A plethora of commercial and open-source tools have emerged that implement inbound oracles.
\emph{Orisi}%
\footnote{%
	Orisi: \url{https://orisi.org/}.
	Provable Things: \url{https://provable.xyz}.
	TinyOracle: \url{https://github.com/axic/tinyoracle}.
	ChainLink: \url{https://chain.link/}.
	All links accessed on June 7, 2020. % \today.
}\addtocounter{footnote}{-1}\addtocounter{Hfootnote}{-1}
is a solution for a distributed set of inbound oracles for Bitcoin, which are executed by independent and trustworthy third parties. 
The majority of all oracles has to agree on the outcome from external data. To fulfill this purpose, money from senders and receivers is parked into a multiple-signature address, including their signatures as well as the signature address of the majority of the oracles result.
In our framework, Orisi is categorized as a pull-based inbound oracle.
\emph{Oraclize}, recently rebranded as \emph{Provable Things}%
\footnotemark %\footnote{Provable Things: \url{https://provable.xyz}. Accessed: \today} %, which works across multiple platforms.
is a popular service for inbound oracles that works with multiple smart-contract-enabled blockchain platforms.
The service acts like a trusted intermediary between blockchains and a variety of independent data sources.
It also provides a mechanism to mitigate \emph{corrupt} oracles~\cite{DBLP:conf/emisa/NeidhardtKN18}. 
Its Provable Engine executes a set of instructions to react as certain conditions are met, thus making it classifiable both as a push-based and a pull-based inbound oracle. 
Other services which are natively classifiable as pull-based follow.
In the Ethereum-specific \emph{TinyOracle}%
\footnotemark 
an intermediary contract acts as a receiver for the actual contract and simultaneously emits an event to the subscribing RPC client.
The \emph{lookup} contract stores both query and respondent addresses, while the \emph{sample client} contract calls the oracle service of TinyOracle.
\emph{Reality Keys} provides a combination of both automated and human-driven pull-based inbound oracles \cite{DBLP:conf/emisa/NeidhardtKN18}.
\emph{Chainlink}%
\footnotemark %\footnote{Chainlink Documentation: \url{https://docs.chain.link/docs}. Accessed: \today}
offers a general-purpose framework for building decentralized inbound oracles, providing decentralization on both oracle and data-source levels. % They use external adapters to manage the communication with an external data source. 
A Chainlink node can have multiple external adapters for different data sources. % and handles individual tasks and jobs with the data and signs transactions to the blockchain. The blockchain node monitors the network for specific events to fulfill data lookups and allows the Chainlink node to broadcast transactions to requesting smart contracts. The Chainlink node subscribes to the Blockchain node and monitors missed blocks if the blockchain node goes offline.
\emph{Witnet} \cite{DBLP:journals/corr/abs-1711-09756} provides a decentralized oracle network protocol based on Ethereum. It also enables miners to earn tokens. An Ethereum bridge is implemented, providing Witnet nodes to run Ethereum nodes with the option to operate with Ether and make contract calls. 

Blockchain inbound oracles have also been considered in a number of research works. 
Xu et al.~\cite{DBLP:conf/wicsa/XuPZGPTC16} introduce the concept of \emph{validation oracles}, 
namely trusted third-party operators (either automatic or human) that act as inbound oracles. 
The authors distinguish between \emph{internal} ones, periodically transmitting external verified data to the blockchain, and \emph{external} ones, operating as trusted external validators of transactions based on information that is external to the blockchain.
According to our scheme, we see that the former is push-based and the latter is pull-based.
Adler et al.~\cite{DBLP:conf/ithings/AdlerBVPVK18} introduce a decentralized pull-based inbound oracle service. The implementation provides a voting game, which decides the truth or inaccuracy of propositions. Players can be \emph{voters} or \emph{certifiers}. While certifiers play a role in cases with the requirement for high accuracy, voters are utilized for low-risk/low-reward roles.  
Due to the random selections of propositions, a level of security is provided against manipulation.
We remark that the successful implementation of random generators is also part of the realization of oracles.  
Zhang et al.~\cite{DBLP:conf/ccs/ZhangCCJS16} present \emph{Town Crier}, a push-based inbound oracle
that acts like a data-feed system connecting a blockchain with a back-end that scrapes HTTPS websites.

We can observe that, thus far, the vast majority of the efforts has been devoted to the design and implementation of inbound oracles. Indeed, a recent technical report of ISO/TC~307 describes oracles for their sole task of providing off-chain information to the blockchain~\cite{ISOTC/2019:ISO/TR2345Blockchainanddistributedledgertechnologies}. In this paper, however, we also investigate and specify the patterns behind the opposite information flow, namely that of outbound oracles, also known as \emph{reverse} oracles~\cite{Xu.etal/EuroPLoP2018:PatternCollectionBlockchainBasedApplications}.

%% file: sections/pattern.tex
\newcommand\patternTableA[5]{%
  {
  \hfill
  \renewcommand*{\arraystretch}{1.0}
  \vspace*{-1.9\baselineskip}
  \begin{table}[h]
    \begin{tabularx}{\linewidth}{@{}|>{\columncolor{gray!20}}X|}
      \hline 
      \textbf{PATTERN: {#1}} \\
      \textbf{Problem} #2 \\
      \textbf{Solution} #3 \\
      \textbf{Benefits} #4 \\
      \textbf{Drawbacks} #5 \\
      \hline
    \end{tabularx}
  \end{table}
  \vspace*{-1.9\baselineskip}
  }%
}
\newcommand\patternTable[5]{%
  {
  \renewcommand*{\arraystretch}{1.0}
  \begin{small}
  \begin{longtable}{|p{\textwidth}|}
  
    \hline
    \rowcolor{gray!40}
    \textbf{PATTERN: {#1}} \\ %\hline 
    \endfirsthead
    
    %\rowcolor{gray!30}
    %continued from previous page \\
    %\textbf{PATTERN: {#1}} \\
    %\endhead

    \rowcolor{gray!20}
    \textbf{Problem} #2 \\
    \rowcolor{gray!20}
    \textbf{Solution} #3 \\
    \rowcolor{gray!20}
    \textbf{Benefits} #4 \\
    \rowcolor{gray!20}
    \textbf{Drawbacks} #5 \\
    \hline
  \end{longtable}
  \end{small}
  }%
}

\begin{figure}[tbp]
	\centering
	\includegraphics[scale=0.7]{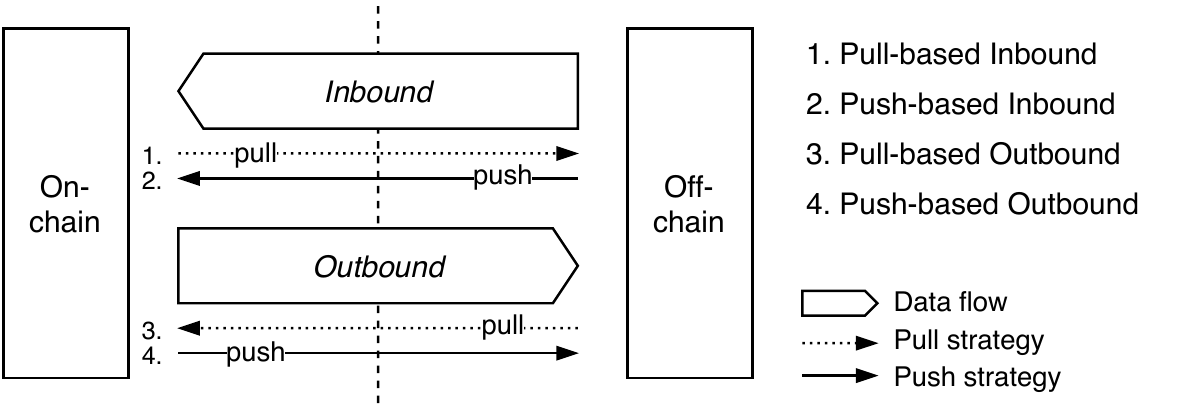}
	\caption{Conceptual overview of the oracle data flow partitioning.}
	\label{fig:oracles:dataflow}
\end{figure}
In this section, we describe in detail basic oracle patterns resulting from the partitioning of the direction (inbound/outbound) and initiation of data flow (pull/push) between on-chain and off-chain components. 
\Cref{fig:oracles:dataflow} shows the data flow along the fundamental dimensions outlined above.
When applying this partitioning, a basic distinction can be made between inbound oracles and outbound oracles, each of which can be further refined according to data pull and push strategies.

\begin{figure}[tbp]
	\centering
	\includegraphics[scale=0.23]{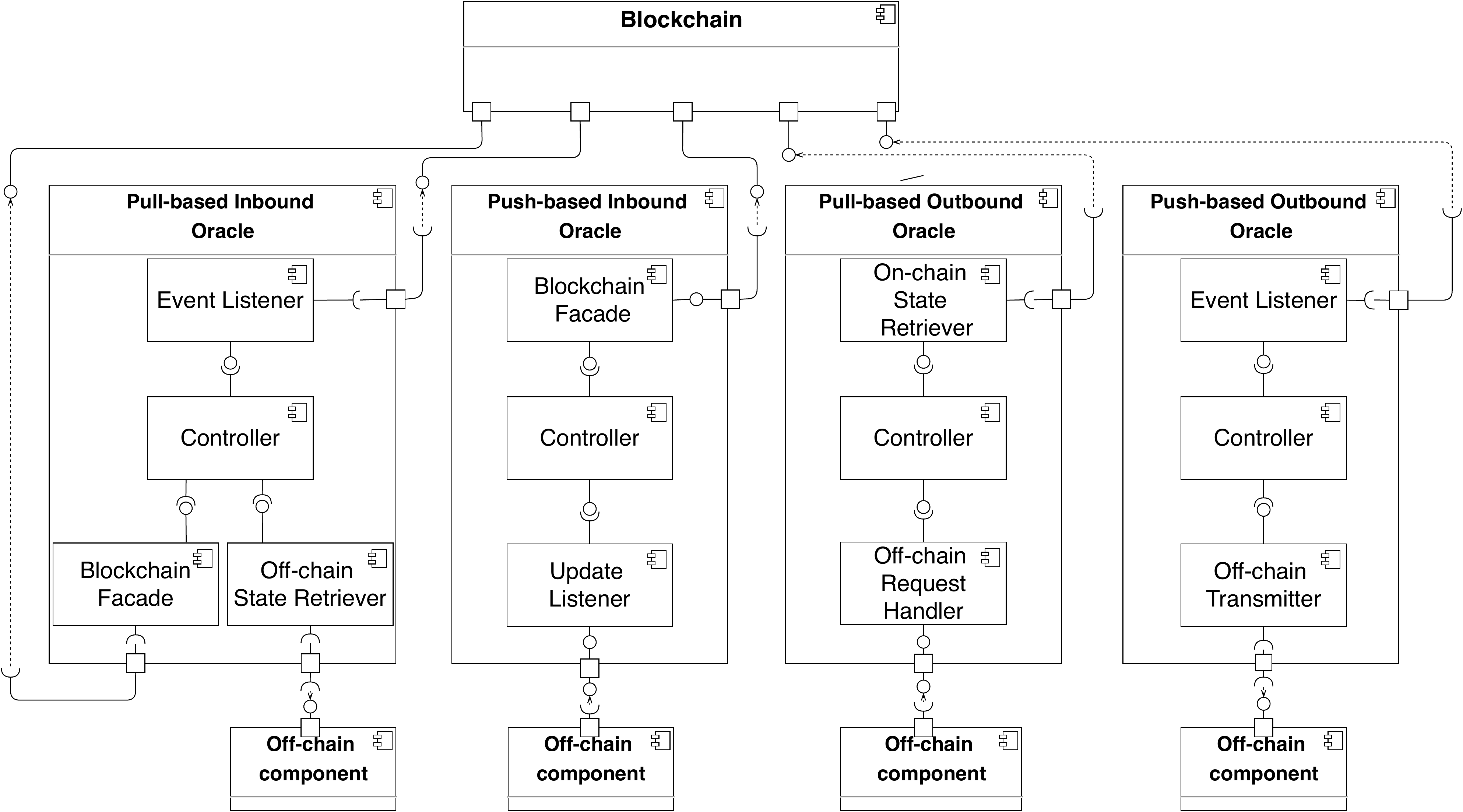}
	\caption{An overview of the oracle types and conceptual structural components.}
	\label{fig:oracles:component}
\end{figure}
%\subsection{Pattern Overview}
Before discussing each pattern in more detail, we first give a general overview of the patterns and their respective conceptual structural components (also called ``pattern participant'') in \cref{fig:oracles:component}. 
The blockchain is considered to be part of a larger software system, with software components being located on and off-chain. In such an environment, it is often necessary to be able to communicate across system boundaries in both directions to exchange information. For example, components on the blockchain (such as smart contracts) may require knowledge from software components outside the blockchain, and vice versa. The outside world requires knowledge from the blockchain, too. 
%The proposed oracle patterns address this issue and act as a middleware between the on-chain and the off-chain world. %, to enable otherwise decoupled software components to exchange data. 
%In other words, they act as a ``service provider'' to enable otherwise decoupled software components to exchange data.
Regarding the terminology used throughout this paper, note that the term ``event'' in relation to the blockchain refers to any activity that can take place on the blockchain (e.g., data is persisted, a transaction occurs, a block is added, etc.).

\subsection{Inbound Oracle}
An inbound oracle transmits information from the outside world to the blockchain. 
As a blockchain cannot directly acquire information from the outside world, it relies on the outside world pushing information into the network. Given this fact, the most obvious approach to obtaining external information on the blockchain is to alert the outside world about the need to push required information into the network. This approach is described in the \pattern{pullinbo} pattern and is characterized by the fact that the exchange of information is initiated on-chain.

%\mysubsubsection{\Gls{pullinbo}}
%\subsubsection{Pull-based Inbound Oracle}
\patternTable{\Gls{pullinbo}}
% PROBLEM
{A blockchain application requires knowledge contained outside of the blockchain, but since blockchains are closed systems, applications cannot directly acquire information from the outside world.}
% SOLUTION
{A \pattern{pullinbo} allows blockchain applications to request states from off-chain components. When a blockchain application requests an off-chain state, the \pattern{pullinbo} receives this request, gathers the state from off-chain components, and sends the result back to the blockchain (via a transaction).}
% BENEFITS
{State requests are initiated in the blockchain. Thus the whole process is transparent. It can be traced whether off-chain data was successfully provided (in time) or not.}
% DRAWBACKS
{State requests have to be initiated from the blockchain, this induces a passive character. Further, the \pattern{pullinbo} response time depends on the speed of the blockchain network, which may lead to a bottleneck. Network congestion may result in delayed or missed off-chain state retrieval, as the oracle only starts working after it registers requests from the blockchain.}
\begin{figure}[tbp]
	\centering
	\includegraphics[scale=0.45]{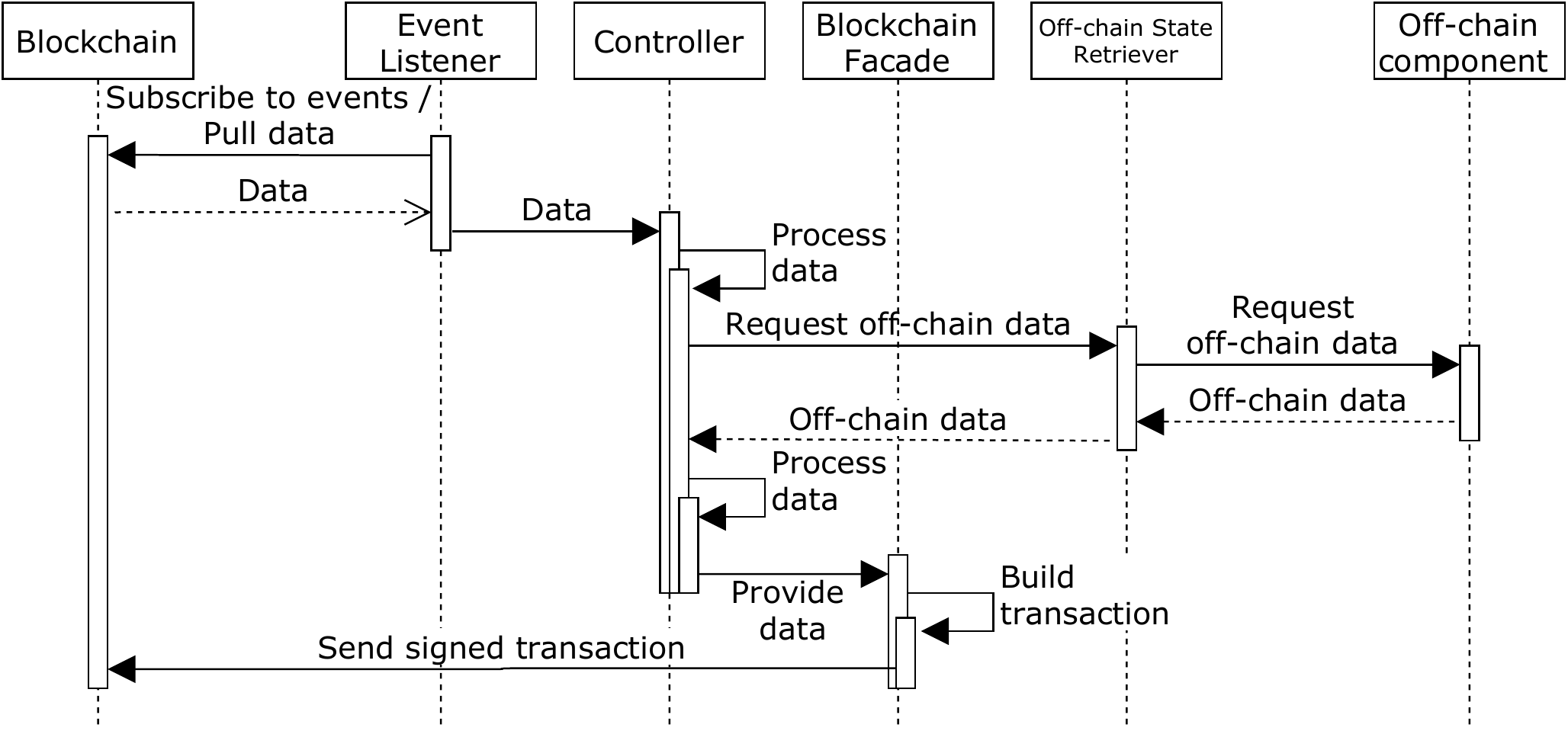}
	\caption{Sequence diagram showing the component interactions for the \pattern{pullinbo}.}
	\label{fig:inbound:pull:sequence}
\end{figure}
\noindent
The conceptual interaction of the pattern participants is shown in \cref{fig:inbound:pull:sequence}: An \textit{Event Listener} subscribes to relevant events on the blockchain, which forwards event data to a \textit{Controller}. The \textit{Controller}  gathers required data from an off-chain component via an \textit{Off-chain State Retriever}. The gathered data may be further processed by the \textit{Controller} before it is returned to the blockchain via a \textit{Blockchain Facade}.

\begin{figure}[bp]
	\centering
	\includegraphics[width=.66\textwidth]{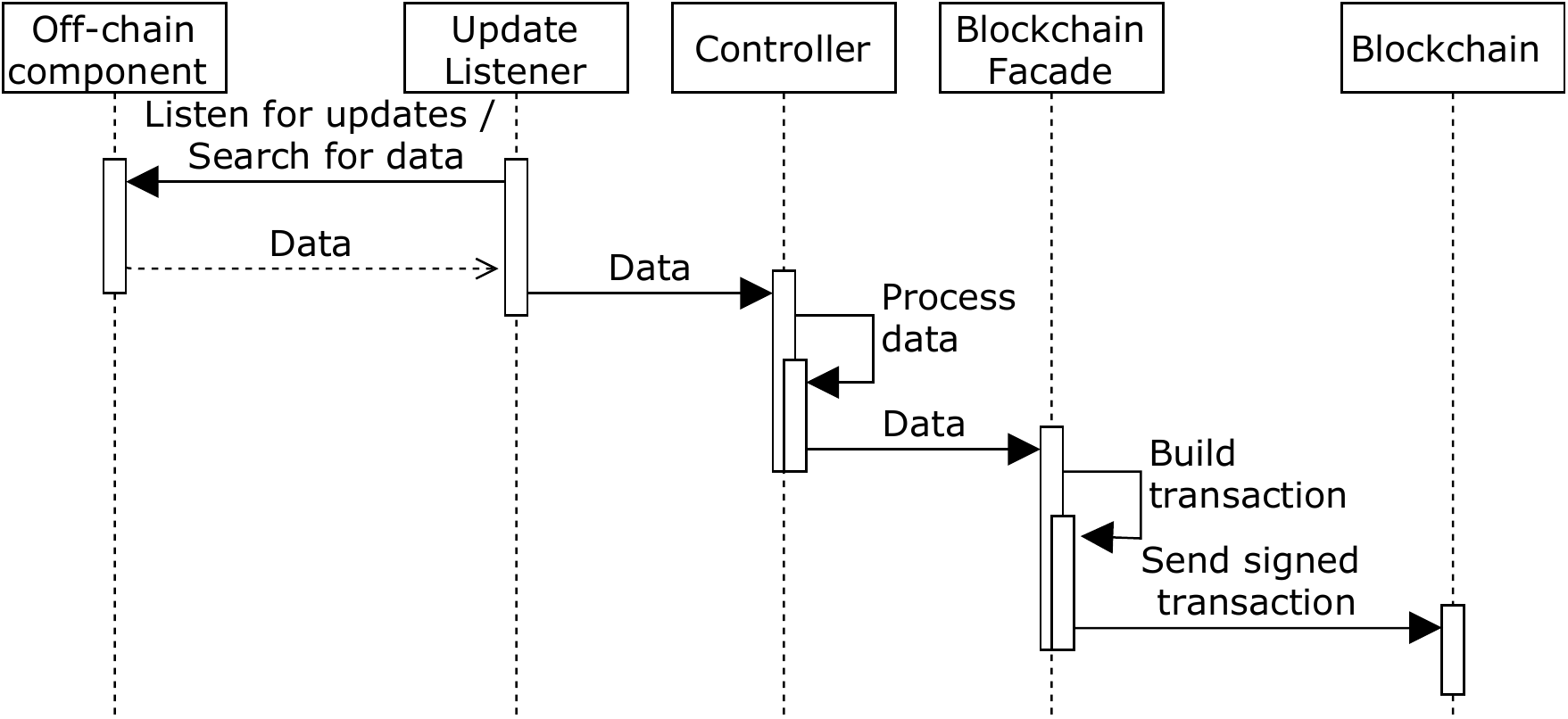}
	\caption{Sequence diagram showing the component interactions for the \pattern{pushinbo}.}
	\label{fig:inbound:push:sequence}
\end{figure}
Another approach to transferring external knowledge to the blockchain is to monitor changes in the off-chain world that are relevant to the blockchain and to transfer these changes to the network. This approach is described by the \pattern{pushinbo} pattern and is characterized by the fact that the exchange of information is initiated off-chain.
%
%\subsubsection{Push-based Inbound Oracle}
\patternTable{\Gls{pushinbo}}
% PROBLEM
{A blockchain application must be supplied with knowledge outside the blockchain, but since blockchains are closed systems, this knowledge cannot be directly communicated.}
% SOLUTION
{A \pattern{pushinbo} allows off-chain information to be propagated to the blockchain by monitoring off-chain state changes and forwarding them to the blockchain.}
% BENEFITS
{Scattered or irregularly updated data outside the blockchain is proactively pushed to the blockchain application. Therefore, the application does not require capabilities to search and query off-chain data. In addition, data can be checked more easily by the \pattern{pushinbo}, considering the limited functionality of blockchain environments.}
% DRAWBACKS
{The \pattern{pushinbo} is not deployed or triggered on the blockchain, making data provision entirely dependent from (non-distributed) applications running off-chain. To manipulate blockchains with incorrect information, an	adversary only needs to compromise the off-chain component(s) from which the oracle receives the data.}

\noindent
The \pattern{pushinbo}, as conceptually illustrated in \cref{fig:inbound:push:sequence}, listens to relevant off-chain component updates via an \textit{Update Listener} and forwards the data to the \textit{Controller}. The \textit{Controller} may process (e.g., filter, verify, etc.) the data before it is sent to the blockchain via a \textit{Blockchain Facade}.

\subsection{Outbound Oracle}
%\subsubsection{Pull-based Outbound Oracle}
%
An outbound oracle transmits information from the blockchain to the outside world. 
Due to its underlying properties, a blockchain can store state information in the form of transactions, but it cannot actively communicate that state to the off-chain world. In light of this, the most obvious path to obtaining data from the blockchain is to fetch it. This approach is described by the \pattern{pulloutbo} pattern and is characterized by the fact that the exchange of information is initiated off-chain.

%\subsubsection{\Gls{pulloutbo}}
\patternTable{\Gls{pulloutbo}}
% PROBLEM
{Knowledge contained on the blockchain is needed  outside the blockchain, but since blockchains are closed systems, the outside world cannot directly request information.}
% SOLUTION
{A \pattern{pulloutbo} allows blockchain data to be queried and filtered to make it available to the outside world. It can be called from (off-chain) components to pull (all) blockchain data and query relevant information.}
% BENEFITS
{The \pattern{pulloutbo} allows to decouple external status requests from the actual status retrieval. Thus, the pattern offers the possibility of uniformly accessing and querying relevant information on the blockchain.}
% DRAWBACKS
{Depending on the size of the blockchain and the knowledge of the location of the requested information, the provision of the data may take some time.}

\begin{figure}[tbp]
	\centering
	\includegraphics[width=.66\textwidth]{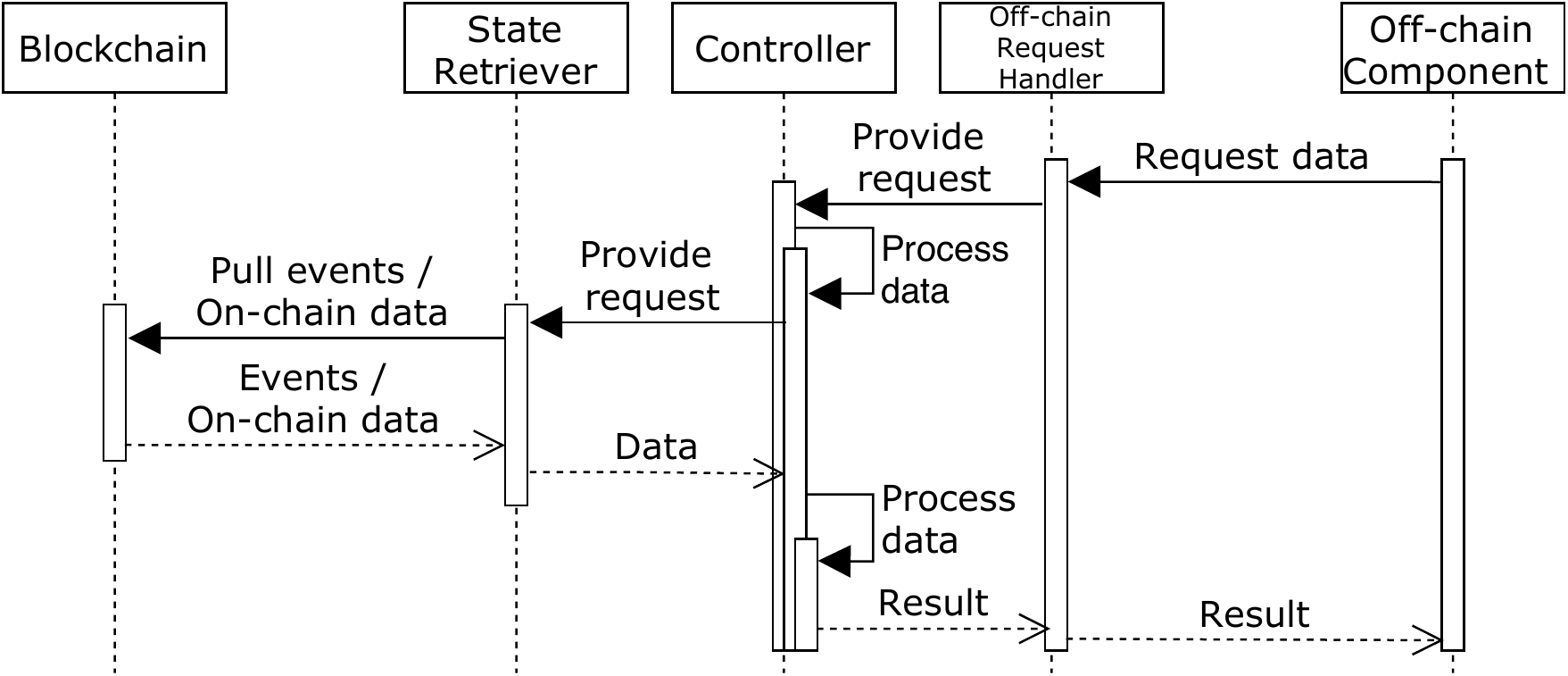}
	\caption{Sequence diagram showing the component interactions for the \pattern{pulloutbo}.}
	\label{fig:outbound:pull:sequence}
\end{figure}
\noindent
The \pattern{pulloutbo}, as conceptually outlined in \cref{fig:outbound:pull:sequence}, receives off-chain data requests via an \textit{Off-chain Request Handler} and forwards the requests to the \textit{Controller} to process the request before forwarding it to the \textit{State Retriever}, which is responsible for retrieving data from the blockchain. The result is returned to the \textit{Controller}, which may process the data before it is sent to the off-chain requester via the \textit{Off-chain Request Handler}.

Another approach to transferring internal information from the blockchain is to observe changes on the blockchain that are relevant to the outside world and to transfer these changes off-chain. This approach is described by the \pattern{pushoutbo} and is characterized by the fact that the exchange of information is initiated on-chain.
%
%\subsubsection{Push-based Outbound Oracle}
\patternTable{\Gls{pushoutbo}}%
% PROBLEM
{Knowledge contained on the blockchain must be available outside the blockchain, but since blockchains are closed systems, applications cannot directly propagate information to the outside world.}
% SOLUTION
{A \pattern{pushoutbo} monitors the blockchain for relevant changes to subsequently trigger or perform activities outside the blockchain.}
% BENEFITS
{The \pattern{pushoutbo} constantly monitors the blockchain. Thus, it is possible to (partially) automate blockchain related tasks by taking action when a blockchain state is updated.}
% DRAWBACKS
{The \pattern{pushoutbo} is required to run continuously in order to monitor all events (on time) on the blockchain. In case the oracle unexpectedly stops, updates (depending on the implementation) may be missed. In addition, depending on the speed of the blockchain network, delays can occur, which can lead to unwanted delays in time-sensitive interactions.}

\begin{figure}[bp]
	\centering
	\includegraphics[scale=0.45]{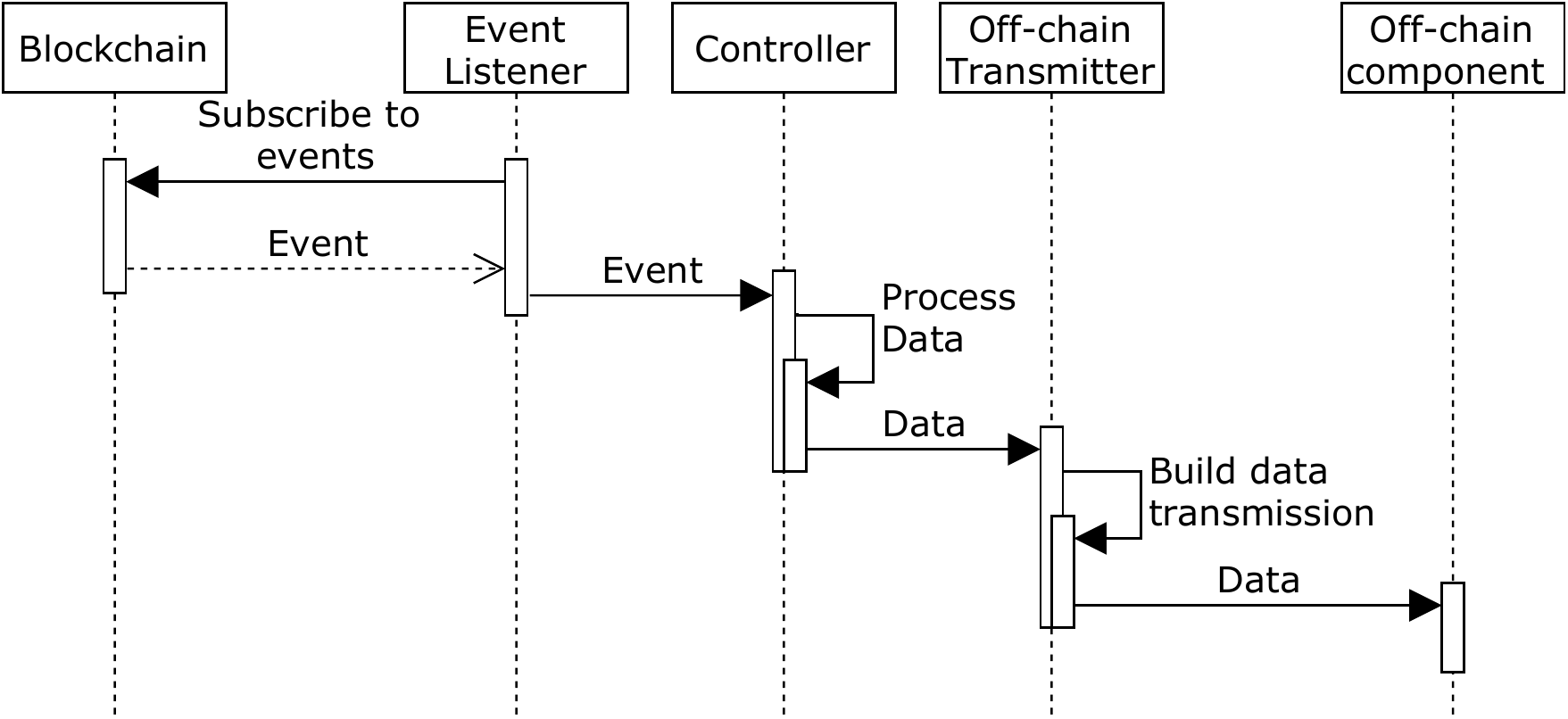}
	\caption{Sequence diagram showing the component interactions for the \pattern{pushoutbo}.}
	\label{fig:outbound:push:sequence}
\end{figure}
\noindent
The \pattern{pushoutbo}, as shown in \cref{fig:outbound:push:sequence}, subscribes to relevant events on the blockchain via an \textit{Event Listener} and forwards event data to the \textit{Controller}, which may process the data before it is sent via the \textit{Off-chain Transmitter} to an off-chain component. 

\pagebreak

%% file: sections/implementation.tex
Among other successful use cases, the blockchain has been adopted as a backbone for the execution of multi-party business processes~\cite{DiCiccio.etal/InfSpektrum2019:BlockchainSupportforCollaborativeBusinessProcesses}. 
This section describes some use cases in that domain we considered to implement the oracle patterns.

\begin{figure}[tbp]
	\centering
	\includegraphics[width=1.0\linewidth]{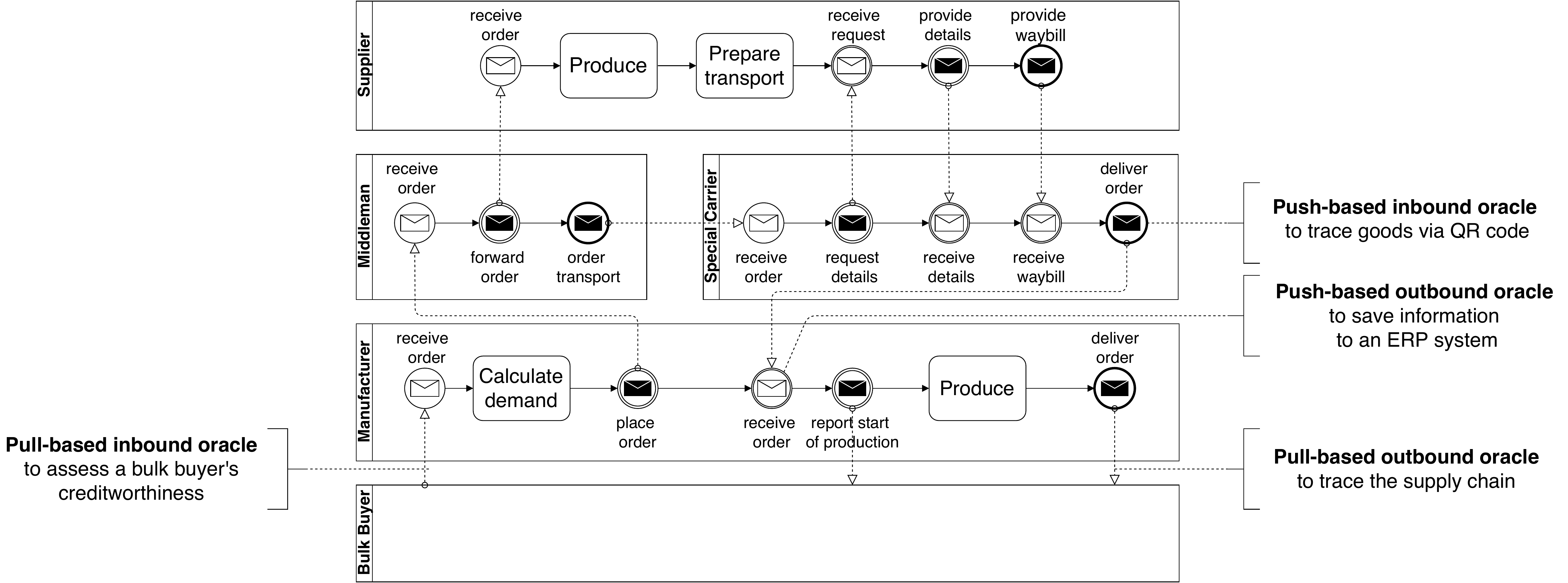}
	\caption{A supply chain process (in BPMN, from \cite{DBLP:conf/bpm/WeberXRGPM16}), showing where oracles are employed.}
	\label{fig:supplychain}
\end{figure}
\Cref{fig:supplychain} illustrates a simplified model of a supply-chain process inspired by \cite{DBLP:conf/bpm/WeberXRGPM16}. The initiator of the process is a bulk buyer who places an order. The order is then forwarded to a manufacturer. The manufacturer, in turn, calculates the needed material and delegates a middleman to forward the order to a supplier and to book the transportation by a special carrier. When materials are ready, the carrier takes care of the transport from the supplier site to the manufacturer's. Finally, the manufacturer produces the goods and delivers them to the bulk buyer.

The execution of the process is tightly bound at many stages to data flows from and toward the blockchain system. The transfer of information from the off-chain world to the on-chain environment and vice-versa is carried out by the oracles. 
We focus in particular on four oracles -- one for each pattern. They are highlighted with textual comments in \cref{fig:supplychain} and detailed next.
Our implementations of those oracles are based on the Ethereum blockchain, \emph{Web3} library and 
\emph{Python}. Our additional modules for QR scans are based on \emph{QR-Code-Scanner}.
\footnote{Web3: \url{https://github.com/ethereum/web3.py}. Python: \url{https://www.python.org/}. QR-Code-Scanner: \url{https://github.com/code-kotis/qr-code-scanner}. All links accessed on June 7, 2020.}
The source code is available, see \cref{fn:code-repo}.

\begin{figure}[tbp]
	\centering
	\includegraphics[width=1.0\textwidth]{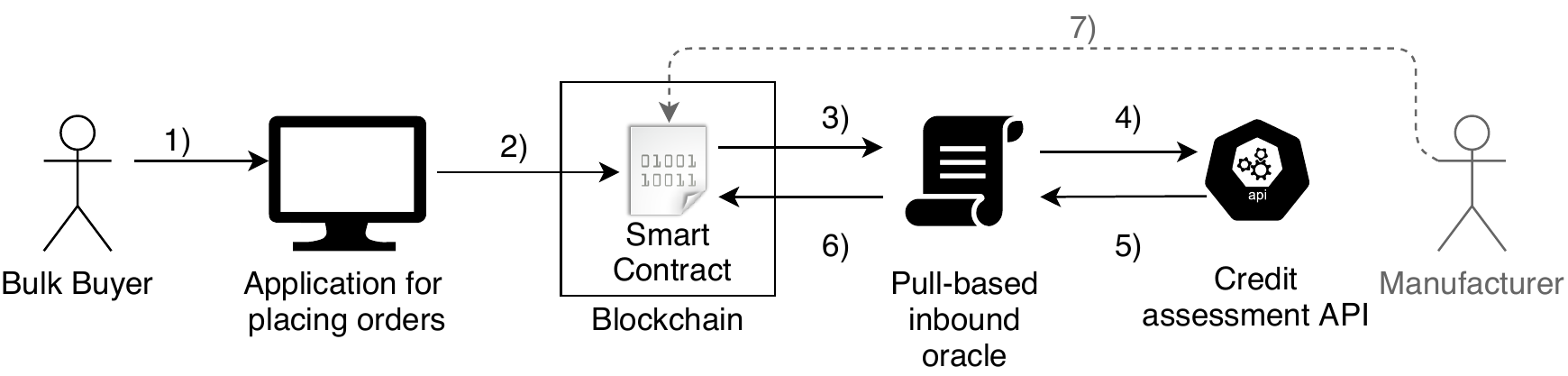}
	\caption[Oracle-based creditworthiness verification]{Oracle-based creditworthiness verification of actors in the supply chain process of \cref{fig:supplychain}.}
	\label{fig:verification}
\end{figure}
\Cref{fig:verification} depicts the oracle-based interaction between a bulk buyer and the manufacturer.
The bulk buyer places an order over a web application (1). The order is forwarded to the manufacturer if the creditworthiness of the buyer is verified.
The order details including the order ID and information on the customer and bulk buyer are forwarded via a transaction to a smart contract (2). The smart contract publishes an event containing information on the bulk buyer such as name and Tax ID.
The \textrm{Event Listener} of a \pattern{pullinbo} subscribes to updates on such events. To retrieve information on the buyer's creditworthiness, the oracle calls the API of an external credit assessment service upon request via its \textrm{Off-chain State Retriever} (4).
As the oracle processes the response (5) with the \textrm{Controller}, it returns this information as transaction data to the smart contract (6) with its \textrm{Blockchain Facade}.
Finally, the manufacturer accesses the order after the verification (7).

\begin{figure}[tbp]
	\centering
	\includegraphics[width=\textwidth]{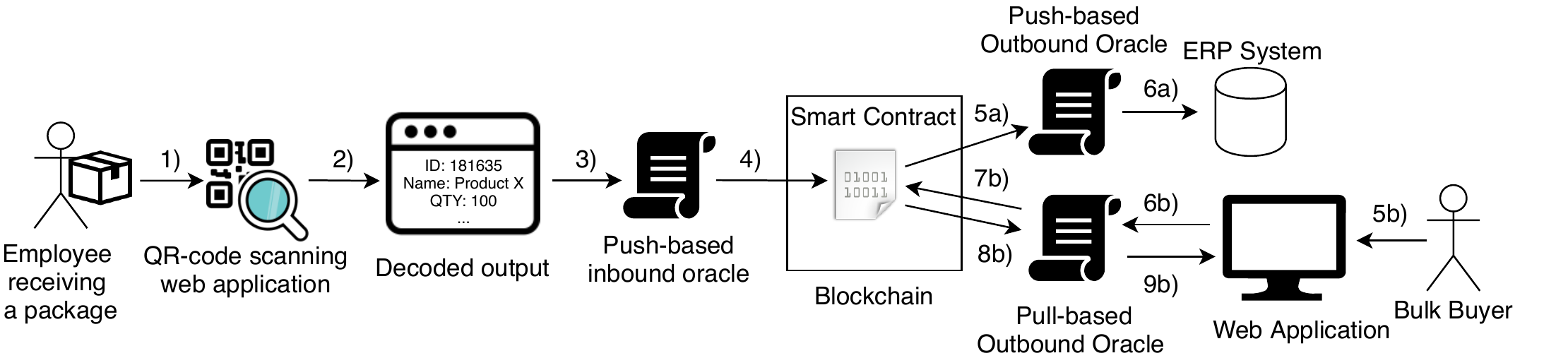}
	\caption[Oracle-based tracing of goods]{Oracle-based tracing of goods via QR Code scanning in the supply chain process of \cref{fig:supplychain}.}
	\label{fig:qr}
\end{figure}
\Cref{fig:qr} illustrates a blockchain-based use-case for the tracing of goods in a supply chain via QR-code scanning. It involves three oracle patterns. The use case starts with an employee registering the delivery of a package. To certify the sending of the package, the employee uses a device with a QR-code scanning application (1). 
The information from the QR code includes the order ID, the name and the quantity of items (2). 
Thereafter, the \pattern{pushinbo} receives the data from the scan (3) via its \textrm{Update Listener}.
The \textrm{Controller} of the oracle encodes the data into a blockchain transaction, enriching it with the location and current timestamp.
Its \textrm{Blockchain Facade} transmits the data to a smart contract (4). The smart contract, in turn, publishes an event that is parsed by the \textrm{Event Listener} of a \pattern{pushoutbo} (5a). 
The \textrm{Controller} of the latter decodes the event data and further passes it along to an ERP system via an \textrm{Off-chain Transmitter} (6a). 
The bulk buyer traces the production of the items identified by the order ID over the blockchain via a web application (5b). Upon request, the web application calls the \textrm{Off-chain Request Handler} of a \pattern{pulloutbo} (6b). 
The oracle \textrm{Controller} turns the request into a query for the \textrm{On-chain State Retriever}. As the requested information is found (8b), the application provides the entire data record on the product(s) back (9b). 
We implemented these use cases to serve as a basis of the analysis described next.

%% file: sections/evaluation.tex
This section describes our findings from a quantitative 
analysis on proof-of-concept implementations of the four oracles, based on the use cases presented above. 

\mysubsubsection{Setup}
We focus on the time and costs dimensions. Regarding time, specifically latency, we are interested in answering two questions. The first question is whether we observe differences in time among the different implemented patterns. This might indicate that dissecting oracles the way we propose in this paper is not only important from a software engineering perspective, but also with respect to the range of use cases they cater for. The second question is whether the observed timings are caused by our experimental settings. We perform all experiments on Ropsten, a test network for Ethereum. We choose Ropsten as it is accepted in the scientific literature for testing purposes \cite{abou2019co,delgado2020blockchain,krejci2020blockchain}. The test code and the code used for the quantitative analysis are available, see \cref{fn:code-repo}.
The smart contract \textit{arrival.sol} mimics the use case from \cref{fig:qr}, which we use to evaluate the \pattern{pushinbo}, the \pattern{pulloutbo} and the \pattern{pushoutbo}. It is deployed at address \texttt{0x1186aEDAb8f37C08CC00a887dBb119787cfE6AAf}.
The smart contract \textit{customer.sol} mimics the use case from \cref{fig:verification}, which we use to evaluate the \pattern{pullinbo}. It is deployed at address \texttt{0x9c2306eccc5afa6ee0c1eca6deab66cc336c3b3d}.

\begin{figure}[tb]
    \centering
    \includegraphics[width=\textwidth]{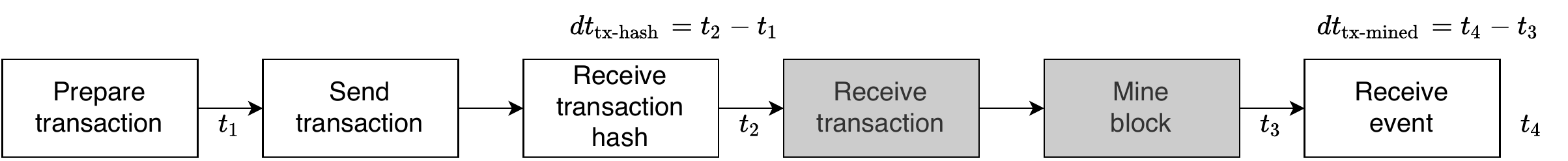}
    \caption{Schematic process for measuring latency, with off-chain (white) and on-chain (grey) tasks.}
    \label{fig:performance:time:explanation}
\end{figure}

To assess the costs of inbound oracles, we measure the consumed gas. 
Note that gas costs also captures the computational and storage effort.
We convert Ether to Euros by using the mean exchange price for Ether over the evaluation period ($144.86$ €/Ether), and gas usage converts to Ether using the gas price of the transactions (on average $ 7.45 \times 10^{-10} $ Ether / gas).

The outbound oracles read from the blockchain and we thus focus on the time dimension.
Note that we keep the retrieval state of the \pattern{pulloutbo} constant to eliminate this as a varying factor. Furthermore, in the implementation of the \pattern{pullinbo} we do not store any states in the receiving smart contract, because the transaction invokes the client smart contract directly and we exclude its handling of the data in the experiment. In contrast, the \pattern{pushinbo} stores the state and emits an event; this is necessary so that the client smart contract can retrieve the state.

To measure latency (see also \cref{fig:performance:time:explanation}) we capture the time between a transaction being sent to the blockchain node ($t_1$) and the time when we receive the transaction hash ($t_2$). We indicate the difference as $dt_{\text{tx-hash}}$. 
For the \pattern{pushoutbo}, we measure the period between the timestamp of the block that included the transaction (i.e., the timestamp when the miner started mining that block, $t_3$), and the time in which we receive the event ($t_4$). We name the difference as $dt_{\text{tx-mined}}$.
When clear from the context, we will refer to both measures as $dt$.
It is debatable whether the mining time should be part of the latency measurement.
Note that the time between the submission of a transaction and its inclusion / commitment on the ledger varies drastically between blockchain platforms. Additionally, various other factors need to be taken into account, such as network congestion and, for commit time on Proof-of-Work blockchains, the number of confirmation blocks which is a user-defined parameter -- see e.g.~\cite{2017-Weber-SRDS} for details and measurements. Here, we measured simple inclusion time without additional confirmation blocks, as a placeholder and to highlight this underlying issue.

\input{tables/qa_summary_statistics_for_each_oracle}

\begin{figure}[tbp]
    \centering
    \includegraphics[width=\textwidth]{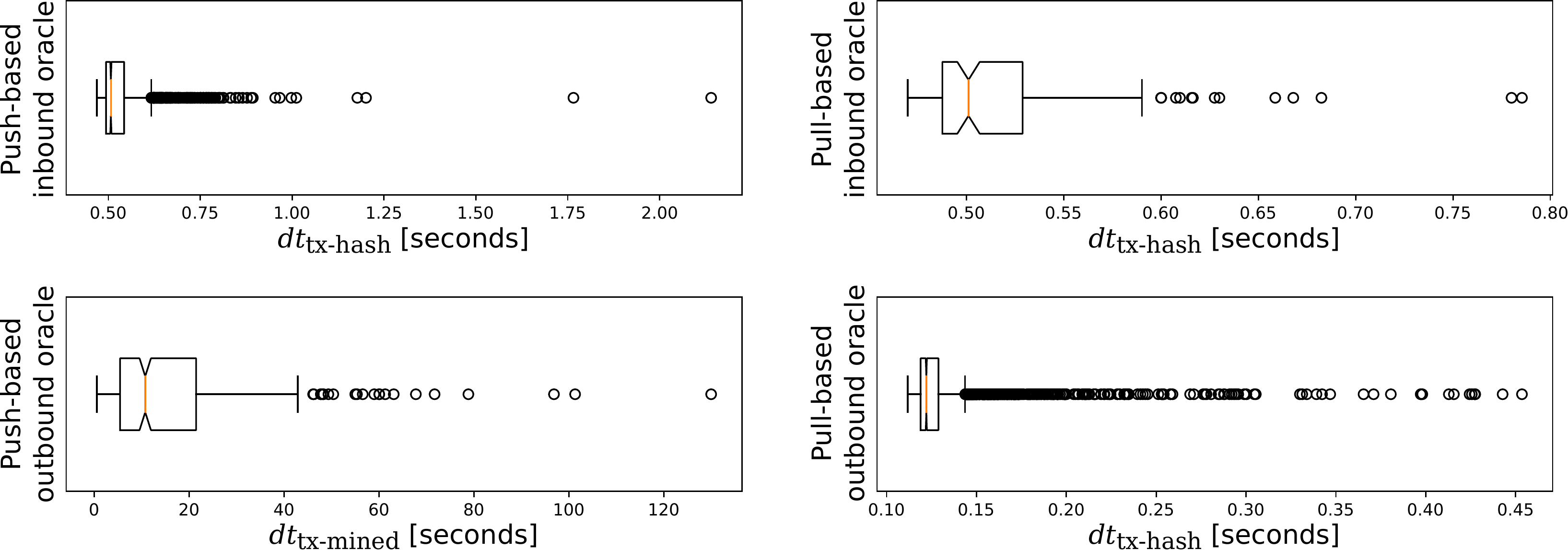}
    \caption{Performance plots for the four oracle implementations.}
    \label{fig:qa:performance:oracles}
\end{figure}

\mysubsubsection{Results}
\Cref{fig:qa:performance:oracles} and \cref{tab:qa:summary:statistics:for:each:oracle} show the results of our experiments.
The \pattern{pulloutbo} is the fastest of the four oracles with a mean $dt$ of $0.13 \pm 0.03$ seconds, while the \pattern{pushoutbo} is the slowest with a mean $dt$ of $16.20 \pm 15.95$ seconds. This difference stems from the fact that the \pattern{pulloutbo} reads historical states from the blockchain, whereas the \pattern{pushoutbo} requires a transaction to be included -- which is subject to high variance and an average delay of roughly 1.5 inter-block times~\cite{Xu.etal/2019:ArchitectureforBlockchainApplications}. 
This transaction triggers the event that is picked up by the \pattern{pushoutbo}.
We received $75\%$ %(i.e., the third quartile) 
of the \pattern{pulloutbo} transactions within $0.12$ seconds. For the \pattern{pushoutbo}, instead, the third quartile amounts to $21.44$ seconds.
From the box plots in \cref{fig:qa:performance:oracles}, we can observe that the $dt$ measurements of the \pattern{pulloutbo} and the \pattern{pushinbo} have a significant number of outliers and follow a long-tail distribution. 
This is less pronounced for the other two oracles.
Discounting outliers, the $dt$ distribution for the \pattern{pullinbo} is similar to \pattern{pushinbo}, with mean $dt$s of $0.52\pm0.05$ and $0.53\pm0.08$, respectively, and the same minimum ($0.46$) and median ($0.50$) values. They differ slightly in their $25$th ($0.48$ vs. $0.49$) and $75$th ($0.52$ vs $0.54$) percentiles.

For \pattern{pushinbo} and \pattern{pullinbo} we measured the transaction costs in Ether, and converted them to Euros with the above-mentioned exchange rate. The results are reported in \cref{tab:qa:summary:statistics:for:each:oracle}.
The gas price setting in our setup relied on the current market price -- which turned out to be highly variable on Ropsten, and not representative of the Ethereum mainnet.
To give an indication of the cost we would have incurred on the mainnet, 
we retrieved the approximate median gas price from the Google BigQuery public database of Ethereum for the period in question, which was $8.5$ Gwei (averaged over $3.15$ million transactions). 
If we multiply this with our mean gas consumption and the exchange rate, we get a median transaction cost of 
$0.028$~€ for  \pattern{pushinbo} and
$0.056$~€ for  \pattern{pullinbo}.

%% file: tables/qa_summary_statistics_for_each_oracle.tex
\begin{table}[tb]
\centering
%% \tablesize{} %% You can specify the fontsize here, e.g., \tablesize{\footnotesize}. If commented out \small will be used.

\caption[Summary statistics of time and costs for oracle invocations.]{Summary statistics of time and costs for oracle invocations (on the Ropsten Ethereum test-net).}
\label{tab:qa:summary:statistics:for:each:oracle}
\def\arraystretch{1.2}
\resizebox{\textwidth}{!}{
\begin{tabular}{@{} l r r r r r r r r @{}}
    \toprule
              & \textbf{$n$} & \textbf{mean} & \textbf{std}	& \textbf{min} & \textbf{$x_{0.25}$} & \textbf{$x_{0.50}$} & \textbf{$x_{0.75}$} & \textbf{max} \\
    % \midrule
    %     \textit{All Oracles} & 296 & & & & & & & \\
    %     \qquad Time Delta [seconds] & & 37.51 & 87.49 & 1.00 & 10.00 & 19.00 & 36.00 & 1,573.00 \\
    %     \qquad Gas Used   & & 53,782.11 & 1,530.16 & 44,202.00 & 54,166.00 & 54,282.00 & 54,310.00 & 54,382.00 \\
    %     \qquad €   &  & 53,782.11 & 1,530.16 & 44,202.00 & 54,166.00 & 54,282.00 & 54,310.00 & 54,382.00 \\
    \midrule
        \textit{Push-based inbound oracle} & 2437 & & & & & & & \\
        \qquad $dt_{\text{tx-hash}}$ [seconds] & & 0.53 & 0.08 & 0.46 & 0.49 & 0.50 & 0.54 & 2.14 \\
        % \qquad $dt_{\text{tx-mined}}$ [seconds] & & 22.81 & 126.97 & -140.01 & 3.10 & 9.08 & 19.11 & 3165.08 \\
        % \qquad Gas Used  & & 44881.86 & 1174.49 & 36883.00 & 45139.00 & 45235.00 & 45259.00 & 45319.00 \\
        % \qquad Transaction Fee [Gwei] & & 14672.81 & 21678.48 & 4.09 & 10.85 & 10.86 & 41047.00 & 117642.20 \\
        % \qquad Transaction Fee [Gwei] & & 35727.01 & 42570.29 & 0.00 & 452.71 & 45151.00 & 45259.00 & 950691.00 \\
        % \qquad Transaction Fee [€]   & & $2.26 \times 10^{-3}$ & $3.35 \times 10^{-3}$ & $6.33 \times 10^{-7}$ & $1.67 \times 10^{-6}$ & $1.68 \times 10^{-6}$ & $6.34 \times 10^{-3}$ & $1.81 \times 10^{-2}$ \\
        % \qquad Transaction Fee [Gas]   & & $44,826.39$ & $1,264.20$ & $36,739.00$ & $45,139.00$ & $45,235.00$ & $45,259.00$ & $45,319.00$ \\
        \qquad Transaction cost [Gas]   & & $44,827$ & $1,265$ & $36,739$ & $45,139$ & $45,235$ & $45,259$ & $45,319$ \\
        % \qquad Transaction Fee [€]   & & $5.52 \times 10^{-3}$ & $6.58 \times 10^{-3}$ & $3.17 \times 10^{-11}$ & $7.00 \times 10^{-5}$ & $6.98 \times 10^{-3}$ & $7.00 \times 10^{-3}$ & $1.47 \times 10^{-1}$ \\ Average
        \qquad Transaction cost [€]   & & $4.96 \times 10^{-3}$ & $5.78 \times 10^{-3}$ & $2.96 \times 10^{-11}$ & $6.55 \times 10^{-5}$ & $6.53 \times 10^{-3}$ & $6.55 \times 10^{-3}$ & $1.37 \times 10^{-1}$ \\
    \midrule
        \textit{Push-based outbound oracle} & 438 & & & & & & & \\
        \qquad $dt_{\text{tx-mined}}$ [seconds] 
        & & 16.20 & 15.95 & 0.53 & 5.41 & 10.71 & 21.44 & 129.95 \\
        % \qquad Gas Used  & & & & & & & & \\
        % \qquad €   & & & & & & & & \\
    \midrule
        \textit{Pull-based inbound oracle} & 126 & & & & & & & \\
        \qquad $dt_{\text{tx-hash}}$ [seconds] & & 0.52 & 0.05 & 0.46 & 0.48 & 0.50 & 0.52 & 0.78 \\
        % \qquad $dt_{\text{tx-mined}}$ [seconds] & & 13.39 & 13.40 & 0.03 & 2.91 & 8.56 & 19.86 & 64.74 \\
        % \qquad Gas Used  & & 22770.0 & 0.0 & 22770.0 & 22770.0 & 22770.0 & 22770.0 & 22770.0 \\
        % \qquad Transaction Fee [Gwei] & & 615.11 & 2737.32 & 5.46 & 5.46 & 5.46 & 5.46 & 12808.12 \\
        % \qquad Transaction Fee [Gas]  & & $22,770.00$ & $0$ & $22,770.00$ & $22,770.00$ & $22,770.00$ & $22,770.00$ & $22,770.00$ \\
        \qquad Transaction cost [Gas]  & & $22,770$ & $0$ & $22,770$ & $22,770$ & $22,770$ & $22,770$ & $22,770$ \\
        % \qquad Transaction Fee [€] & & $1.44\times10^{-3}$ & $1.53\times10^{-3}$ & $6.33\times10^{-7}$ & $1.67\times10^{-6}$ & $2.09\times10^{-3}$ & $2.10\times10^{-3}$ & $3.94\times10^{-3}$ \\ Average
        \qquad Transaction cost [€] & & $8.91\times10^{-5}$ & $3.96\times10^{-4}$ & $7.91\times10^{-7}$ & $7.91\times10^{-7}$ & $7.91\times10^{-7}$ & $7.91\times10^{-7}$ & $1.85\times10^{-3}$ \\ % Start Price
    \midrule
        \textit{Pull-based outbound oracle} & 2611 & & & & & & & \\
        \qquad $dt_{\text{tx-hash}}$ [seconds] & & 0.13 & 0.03 & 0.11 & 0.11 & 0.12 & 0.12 & 0.45 \\
        % \qquad Gas Used  & & & & & & & & \\
        % \qquad €   & & & & & & & & \\
    % \midrule 
    %      \textit{Round Trip} & 438 & & & & & & & \\
    %      \qquad $dt_{round\text{ }trip}$ [seconds] & & 34.32 & 40.26 & 2.82 & 14.96 & 26.11 & 44.21 & 693.71 \\
    \bottomrule
\end{tabular}
}
\end{table}

%% file: sections/discussion.tex
In the following, we discuss advantages and disadvantages, our experience from the implementation process,
the results analysis above, and finally the limitations and threats to validity of this work.
An advantage of the foundational viewpoint taken in this paper is the clear separation and composition of concerns we can achieve. For example, our implementation, following the patterns in this paper, enables us to implement logic for distinct abstraction levels. As such, it is possible to implement behaviour for all oracles. % \pattern{pushinbo}s. 
More crucially, adding or changing information sources % state variable 
to the oracle only requires us to revise the sole oracle % another class 
without the need to change the on-chain implementation logic. 

Regarding the results of the analysis, we find that latency and cost are both not particularly high.
For instance, when comparing the latency with results from \cite{2017-Weber-SRDS}, where the median commit time of transactions was around $200$ seconds, it is fair to say that the sub-second latency measured in almost all cases (where no transaction inclusion time is part of the latency) is relatively low. This, however, may be different if other blockchain platforms or consensus algorithms are used.

As for cost, we found that a single interaction of either inbound oracle did not incur high fees.
For the fairest possible comparison, gas consumption should be used as a metric as it does not depend on current market prices. Comparing the results on this basis, in \cite{2017-Garcia-BPM} (a cost-optimized version of \cite{DBLP:conf/bpm/WeberXRGPM16}) transactions have a typical gas consumption of 24,000 to 27,000 gas.
This is in line with the \pattern{pullinbo}'s gas consumption; for the \pattern{pushinbo}'s gas usage the additional storage cost accounts for the higher gas cost.
Specific implementations of this pattern can be optimized in this regard, in particular by storing data on-chain only when necessary.
This may be particularly important when many oracle invocations are expected in a given setting, and cost and time delays would add up.

The work we present in this paper has a number of limitations and threats to validity. The patterns are mined using a qualitative mining process (as it is usual). Thus, possible misinterpretations or biases of individual researchers or the whole author team cannot be fully excluded and might have influenced our results. 
Generalizability can only be claimed for the studied technologies (see \cref{sec:background}), but we aimed to define foundational patterns to mitigate this threat as far as possible. Therefore, despite our implementation resorts on Ethereum, our findings are applicable to other blockhain platforms. Nevertheless, we do not claim any form of completeness. Our analyses are preliminary and can only provide a rough indication of time performance and costs; for claiming generalizability beyond the scope of the studied cases, more research would be needed.
Furthermore, the use of a testnet like Ropsten may reduce the representativeness of the analysis results for practical applications. We mitigated these effects by not relying on time and cost measurements from the testnet in our discussion, and by basing relevant cost analyses on data from the Ethereum mainnet instead.
In future work, we will also study different strategies on data structures and message rates to further mitigate the impact that information exchanges have on the overall execution costs.

%% file: sections/conclusion.tex
In this paper, we have investigated how blockchain oracles can be characterized for the communication between the on-chain and off-chain realms.
We abstract individual technical solutions adopted in existing implementations into four foundational oracle patterns. In addition, we have studied their relations, benefits, liabilities, and consequences. 
Finally, we have quantitatively analysed the four patterns in terms of time performance (latency) and cost impacts. 
We find that neither cost nor latency are particularly high for a single invocation of any of the patterns, except that latency can be dominated by transaction inclusion time.
Also, in our experiments the patterns were in most cases subject to different distributions in terms of cost and latency; the results show these characteristic differences.

In future research, we will deepen our analysis with further studies conducted on multiple blockchain platforms, further study how exchanged data rate and quantity has an impact on execution costs, and apply the patterns to more use cases spanning over different fields including autonomous robotic swarm systems~\cite{strobel2020blockchain}.
Furthermore, we want to study the use of patterns for information exchange between blockchains.
The combination of oracle patterns would also be the subject of our future studies.

%% file: sections/acknowledgements.tex
\mysubsubsection{Acknowledgements}
The authors want to thank the Research Institute for Computational Methods at WU Vienna for supplying computational resources. E.\ Castell{\'o} Ferrer
acknowledges support from the Marie Skłodowska-Curie actions (EU project BROS—DLV-751615).
The work of C.\ Di Ciccio was partly supported by MIUR under grant ``Dipartimenti di eccellenza 2018-2022'' of the Department of Computer Science at Sapienza University of Rome.